\documentclass[reprint,superscriptaddress,amsmath,amssymb,aps,prl,]{revtex4-2}

\usepackage{graphicx}
\usepackage{dcolumn}
\usepackage{bm}
\usepackage{bookmark}
\usepackage{xcolor}
\usepackage{lmodern}
\usepackage{upquote}
\usepackage{microtype}
\usepackage{xurl}
\usepackage{hyperref}
\usepackage{svg}
\hypersetup{
  colorlinks=true,
  linkcolor=blue,
  citecolor=blue,
  urlcolor=blue,
  pdftitle={Conditional probability density functional theory for solids},
  pdfauthor={Peiwei You et al.}
}

\newcommand{\huji}{Fritz Haber Center for Molecular Dynamics, Institute of Chemistry, The Hebrew University of Jerusalem, Jerusalem 91904, Israel}
\newcommand{\uci}{Departments of Physics and Astronomy and of Chemistry, University of California, Irvine, California 92697, USA}
\newcommand{\tias}{Quantum Dynamics Laboratory, Tsientang Institute of Advanced Study, Hangzhou, Zhejiang, 310024, China}

\begin{document}

\title{Conditional probability density functional theory for solids}

\author{Peiwei You}
\email{peiwei.you@mail.huji.ac.il}
\affiliation{\huji}
\affiliation{Center for Quantum Matter, School of Physics, Zhejiang University, Hangzhou 310058, China}

\author{Ryan Pederson}
\affiliation{\uci}

\author{Kieron Burke}
\email{kieron@uci.edu}
\affiliation{\uci}

\author{E. K. U. Gross}
\email{ekugross@tias.ac.cn}
\affiliation{\tias}
\affiliation{\huji}

\begin{abstract}
The recently developed conditional probability density functional theory (CP-DFT) yields direct access to the exchange-correlation hole of a system, an important correlation function that is not available in standard DFT. We present the first results for extended materials with periodic boundary conditions. We validate CP-DFT for weakly correlated Na and Si. We apply it to a stretched hydrogen chain, the prototype of strong correlation. When applied to the Kagome material CsV$_3$Sb$_5$, we find $d$-orbital correlations that are missing in standard DFT, implying an enhanced finding probability between two separated electrons and a local charge density wave related modulation.
\end{abstract}

\maketitle

Kohn-Sham (KS) density functional theory (DFT) is the cornerstone for modeling complex chemical and material systems~\cite{Jones2015}. Its widespread utility is due to the development of exchange-correlation (XC) functionals that, despite being approximations, reliably predict ground-state energies and structural properties~\cite{Lejaeghere01012014}. While these functionals are often designed to satisfy exact physical conditions~\cite{Kaplan2023} and yield usefully accurate energetics, they typically rely on a simplified mapping that obscures the intricate, real-space electron correlations that are central to modern condensed matter physics.

One primary limitation involves the representation of the pair density, which characterizes the probability of finding an electron at a specific coordinate relative to a reference electron at another position~\cite{BurkePerdew1995}. This quantity encompasses the XC hole, the depletion of electron density surrounding a reference electron arising from Pauli exclusion and Coulomb repulsion. In conventional KS-DFT, the complex spatial topology of this hole is typically reduced to simplified "on-top" or spherically averaged approximations~\cite{PerdewSavin1995,GoriGiorgi2005,GoriGiorgi2007,LiManni2014}. While these models are effective for determining total energies, they purposely obscure the pointwise fine structure of the electronic environment~\cite{JonesGunnarsson1989}, in favor of system- and spherical-averages that yield XC energies. This loss of spatial resolution could be critical because detailed correlation structures can be the fundamental physical drivers of emergent phenomena. In Mott insulators such as NiO and MnO, for example,
strong on-site Coulomb repulsion drives the formation of the Mott gap \cite{Agapito2015,Sharma2013}. While charge density wave (CDW) formation is often attributed to momentum-dependent electron-phonon coupling and Fermi surface nesting \cite{Zhu2015}, an emerging perspective highlights the potential involvement of electron correlations in CDW and even pair density wave orders in Kagome metals upon entering the superconducting state \cite{Chen2021,Tan2021,Song2023}. Determining the pair density to uncover the nature of these correlations in such electronic orders represents an important step toward a deeper understanding of correlated quantum matter.

Recently, conditional probability (CP) density functional theory has been proposed~\cite{McCarty2020} as an alternative to KS-DFT, especially for systems with strong correlations~\cite{Pederson2022}. CP-DFT is a formally exact framework to directly calculate pointwise pair densities, offering a way to capture the correlated structure of the electronic environment~\cite{McCarty2020}. In CP-DFT, the pair density itself is directly calculated by a sequence of Kohn-Sham-like calculations at every real space position $\mathbf{r}$ in the system. These calculations employ an approximate CP-KS potential to yield the CP density, $n_{\mathbf{r}}(\mathbf{r}')$, the conditional probability of finding an electron at $\mathbf{r}'$ given an electron at $\bf r$, and thus the pair density $P(\mathbf{r}, \mathbf{r'}) = n(\mathbf{r}) n_{\mathbf{r}}(\mathbf{r}')$, where $n(\mathbf{r})$ is the ground-state electron density. Since the pair density determines the interaction and total energies, CP-DFT bypasses the need to capture strong correlation effects in an explicit density functional. Within this paradigm, CP-DFT correctly dissociates H$_2$ and arbitrary-length H chains in the large-separation limit, a classic example where KS-DFT XC functionals notoriously fail~\cite{McCarty2020,Pederson2022}.

Previous applications of CP-DFT were limited to atomic and molecular systems~\cite{Pederson2022} and simple models~\cite{Perchak2022} where standard KS-DFT typically fails. Our work extends CP-DFT to periodic systems, establishing a highly scalable framework for solid-state materials. By incorporating periodic boundary conditions and leveraging parallelization, direct calculation of the pair density becomes feasible for realistic crystalline systems of interest.

We implement this first-principles approach by solving the CP-KS equations self-consistently for each reference point across the periodic cell. We validate using both isolated (Helium) and periodic (Sodium, Silicon) systems, resolving the pointwise structure of the XC hole in a crystalline environment. For strong correlation, we use long stretched hydrogen chains~\cite{Motta2020Hydrogen}, finding an XC hole with power-law decay comparable to accurate many-body correlation benchmarks. For complex quantum matter, we apply CP-DFT to the Kagome metal CsV$_3$Sb$_5$, finding spatial modulations of the pair density and $d$-orbital correlated structures. These correlations lead to a local CDW-related modulation and improved density of states near the Fermi level. None of these structures appear in standard DFT models of holes, showing how calculation of the pair density provides a deeper understanding of the electronic structure in strongly correlated materials.

Consider an $N$-electron system subject to a scaled Coulomb repulsion, characterized by a non-negative coupling strength $\lambda$, such that the ground-state electronic density $n(\mathbf{r})$ remains fixed~\cite{LangrethPerdew1975,Fiolhais2003}. The $\lambda$-dependent pair density is $P^{\lambda}(\mathbf{r},\mathbf{r}') = n(\mathbf{r})n_{\mathbf{r}}^{\lambda}(\mathbf{r}')$, where $n_{\mathbf{r}}^{\lambda}(\mathbf{r}')$ is the CP density, representing the probability of finding an electron at $\mathbf{r}'$ given a reference electron located at $\mathbf{r}$ and a Coulomb repulsion scaled by $\lambda$. CP-DFT determines this pointwise $\lambda$-dependent density by self-consistently solving the CP-KS equations for an $(N-1)$-electron system with an effective potential $v_{\text{s},\mathbf{r}}^{\lambda}\left( \mathbf{r}' \right)$:
\begin{equation}
\begin{aligned}
v_{\mathrm{s},\mathbf r}^{\lambda}(\mathbf r')
=&\ v_{\mathrm{s}}[n](\mathbf r')
- v_{\mathrm{HXC}}^{\lambda}[n](\mathbf r')
+ \Delta v_{\mathbf r}^{\lambda}(\mathbf r') \\
&+ v_{\mathrm{HXC}}^{\lambda}[n_{\mathbf r}^{\lambda}](\mathbf r'),
\end{aligned}
\label{eq:cp_ks_potential}
\end{equation}
where $v_{\text{s}}[n](\mathbf{r'})$ is the standard KS potential for the ground-state density. The terms $v_{\text{HXC}}^{\lambda}[n](\mathbf{r}')$ and $v_{\text{HXC}}^{\lambda}[ n_{\mathbf{r}}^{\lambda}](\mathbf{r}')$ are the $\lambda$-dependent Hartree and XC components evaluated on the ground-state and CP densities, respectively, while $\Delta v_{\mathbf{r}}^{\lambda}\left( \mathbf{r}' \right)$ serves as the CP correction potential~\cite{Pederson2022} (Supplementary Note 1 \cite{SuppMat}). For periodic systems, the CP-KS equation is represented in a numerical atomic-orbital (NAO) basis at each $\mathbf{k}$ point. The reference-dependent CP-KS Hamiltonian acting on the coordinate $\mathbf r'$ is $\hat H_{\mathrm{CP},\mathbf r}^{\lambda}=-\frac{1}{2}\nabla_{\mathbf r'}^2+v_{\mathrm{s},\mathbf r}^{\lambda}(\mathbf r')$, with periodic matrix
\begin{equation}
\begin{aligned}
H_{i\alpha,j\beta,\mathbf k}^{\mathbf r,\lambda}
=&\sum_s e^{-i\mathbf k\cdot\mathbf R_s}
\Big\langle \xi_{i\alpha}(\mathbf r'+\mathbf R_s+\mathbf b_i)
\Big| \\
&\quad \hat H_{\mathrm{CP},\mathbf r}^{\lambda}
\Big| \xi_{j\beta}(\mathbf r'+\mathbf b_j) \Big\rangle,
\end{aligned}
\end{equation}
where $\xi_{i\alpha}$ is the $\alpha$-th NAO associated with atom $i$, $\mathbf b_i$ is the atomic position, and $\mathbf R_s$ is a lattice vector. The CP-KS states are from the generalized eigenvalue equation
\begin{equation}
H_{\mathbf k}^{\mathbf r,\lambda}
C_{n\mathbf k}^{\mathbf r,\lambda}
=\varepsilon_{n\mathbf k}^{\mathbf r,\lambda}
S_{\mathbf k}C_{n\mathbf k}^{\mathbf r,\lambda}.
\end{equation}
Here $S_{\mathbf k}$ is the overlap matrix, $C_{n\mathbf k}^{\mathbf r,\lambda}$ is the eigenvector of the $n$-th CP-KS state at $\mathbf k$, and $\varepsilon_{n\mathbf k}^{\mathbf r,\lambda}$ is the corresponding eigenvalue.
CP-DFT is formally exact, yielding the exact CP density given the exact XC and CP potentials, but these must be approximated in practical applications. We employ traditional KS-DFT XC functionals to obtain the necessary XC potentials, using the Perdew–Burke–Ernzerhof (PBE) generalized gradient approximation~\cite{Perdew1996} throughout this study. To approximate the CP correction potential, we use the original methodology~\cite{McCarty2020,Pederson2022}. This classically inspired approximation satisfies several key exact conditions within CP-DFT~\cite{GoriGiorgi2005,Pederson2022}. 

\begin{figure}[htbp]
\centering
\includegraphics[width=0.47\textwidth]{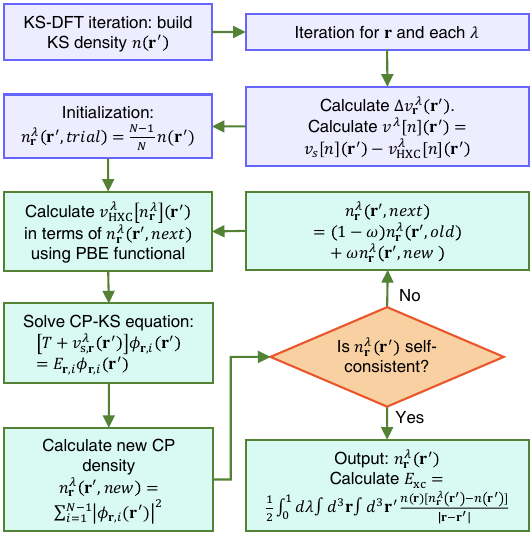}
\caption{Schematic flowchart of the CP-DFT framework. A standard KS-DFT is first performed to find the KS density and potential. To obtain the CP density, electronic iterations are performed for reference points and each coupling strength.}
\label{fig:scheme}
\end{figure}

Computationally, the CP-KS equations use only standard KS-DFT methodology. We employ the SIESTA code~\cite{Soler2002,Garcia2020}, using its NAO basis to handle large systems with extensive real-space grids, and pseudopotentials to avoid calculating core electrons, obtaining the pseudo-density in CP-DFT as in Fig.~\ref{fig:scheme}. The $\lambda$-dependent Hartree-XC potentials are calculated via uniform coordinate scaling relations~\cite{Gorling1993} (Supplementary Note 1 \cite{SuppMat}). For many-electron systems, we use the short-range Gaussian repulsion introduced in earlier work~\cite{Perchak2022} (Supplementary Note 2~\cite{SuppMat}).

We first validate our methodology on weakly correlated systems: bulk Na and Si, and an isolated He atom.
In the exchange limit ($\lambda=0$), the CP density at the reference point satisfies $n_{\mathbf{r}}^{\lambda=0}(\mathbf{r})=n(\mathbf{r})/2$, as confirmed for He and Na (Supplementary Figs.~S1--S4~\cite{SuppMat}). In Na, the exchange hole follows the KS density, while Coulomb correlation reshapes the XC hole, illustrating the generic picture for weakly correlated systems~\cite{Burke1998}. The averaged Na XC hole is featureless and close to LDA (Supplementary Fig.~S5~\cite{SuppMat}), while bulk Si shows strong repulsion in bonding regions (Supplementary Fig.~S6~\cite{SuppMat}). The XC energies from CP-DFT, integrated over all real-space points, agree with KS-PBE values: $-8.4$ vs. $-8.16$ eV for Na, $-28.3$ vs. $-28.11$ eV for He, and $-65.5$ vs. $-64.89$ eV for Si, demonstrating consistency with PBE for weakly correlated systems.

\begin{figure}[htbp]
\centering
\includegraphics[width=0.47\textwidth]{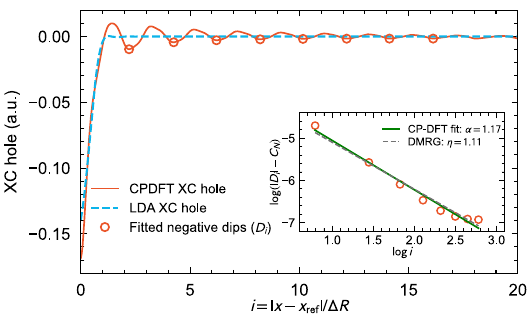}
\caption{
Right-side XC-hole response of the periodic $N=100$ H chain at $\Delta R=2.0$ Bohr with the reference electron at the center. Red circles mark the negative minima $D(i)$. The inset shows $\lvert D(i)\rvert$ versus relative site on logarithmic scales after subtracting the finite-size background, compared with the DMRG spin-correlation exponent $\eta\simeq1.1$ from Motta \textit{et al.}~\cite{Motta2020Hydrogen}.
}
\label{fig:hchain}
\end{figure}

Turning to strong correlation, the prototype is a stretched chain of equally-spaced hydrogen atoms, which is a Mott-Hubbard insulator, but whose exact KS system is a metal~\cite{Stoudenmire2012}. This has been used to validate and compare strong-correlation methods such as DMRG, DMFT, and various flavors of QMC~\cite{Motta2017,Motta2020Hydrogen}, allowing comparison with accurate benchmarks.
Fig.~\ref{fig:hchain} shows the CP-DFT XC hole of a periodic 100-atom chain at the separation of $2.0$ Bohr, with the LDA hole for comparison.
Similar oscillatory charge correlations are reported at comparable separations in ab initio FT-AFQMC~\cite{Liu2020Hydrogen}. Here, the spin-summed CP-DFT XC hole further reveals a long-range power-law decay with $\alpha=1.17$, close to the published spin-correlation exponent $\eta\simeq1.1$~\cite{Motta2020Hydrogen}, providing evidence of long-range correlation in the spin-symmetric regime.
For open chains, the XC-hole dip scales with chain length as $d=0.58$ (Supplementary Fig.~S7~\cite{SuppMat}), close to the density-dimerization exponent of about $0.6$~\cite{Motta2020Hydrogen}.
No semilocal DFT, starting from the uniform gas, could produce such long-range behavior. As the separation is increased, CP-DFT recovers the exact dissociation limit~\cite{Pederson2022}.

\begin{figure}[htbp]
\centering
\includegraphics[width=0.5\textwidth]{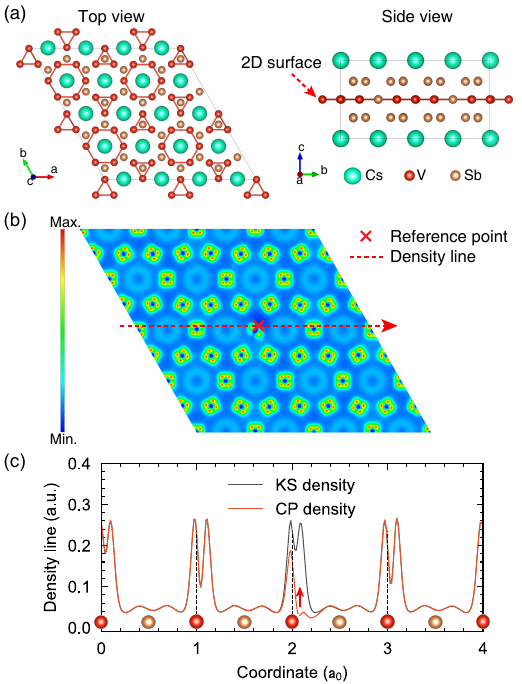}
\caption{(a) Top and side view of CsV$_3$Sb$_5$, with ISD (144 atoms in a 4$\times$4$\times$1 supercell). Green, red and brown balls indicate Cs, V and Sb atoms, respectively. (b) The 4$\times$4 CP density in the 2D surface shown in (a), which is also the value of pair density scaled by $1/n\left( \mathbf{r} \right)$ at the reference point (red cross). (c) CP density and the KS density, along the specific line shown in (b) ($a_0$ is the single cell lattice constant).}
\label{fig:Kagome}
\end{figure}

Last, we apply CP-DFT to the much more complex Kagome metal CsV$_3$Sb$_5$, which provides an excellent proving ground for studying nontrivial band topology~\cite{Ortiz2020}, CDW, and superconductivity~\cite{Liang2021,Zhao2021}. At room temperature, CsV$_3$Sb$_5$ maintains a hexagonal structure comprising a V-atom Kagome layer sandwiched by honeycomb layers of Sb atoms (Fig.~\ref{fig:Kagome}a). Its electronic band structure features van Hove singularities and Dirac points near the Fermi surface. Upon cooling, CsV$_3$Sb$_5$ undergoes a CDW transition at $80$--$100$~K, followed by a superconducting transition below $\sim 2.8$~K~\cite{Chen2021}. The system exhibits $2\times 2$ CDW phases characterized by "star of David" structural distortions from the pristine Kagome lattice, as well as an inverse displacement phase. Theoretical work suggests the most stable configuration favors this inverse star of David (ISD) deformation~\cite{Tan2021}, where the V atoms form the distorted structure shown in Fig.~\ref{fig:Kagome}a. To account for long-range electronic correlations, we employ a $4\times 4\times 1$ supercell, sampling the Brillouin zone with a $2\times 2\times 2$ K-point mesh (detailed in Supplementary Note~2~\cite{SuppMat}). Our calculations utilize supercell lattice constants of $a=21.98$~\AA\ and $c=9.88$~\AA, an energy cutoff of $600$~Ry, and $12$ million real-space mesh points. 

In Fig.~\ref{fig:Kagome}b, we present the pointwise CP density at the V-Sb layer exhibiting the ISD distortion. As paired electrons are observed near the V atoms~\cite{Chen2021}, we choose our reference point within a high-density region near a Vanadium atom ( at $x=2.083~a_0$, $y=2.083~a_0$, beyond the core) to explicitly probe the localized electron interactions. As shown by the local density of states (Supplementary Fig. S8~\cite{SuppMat}), the density here is dominated by V $3d$ orbitals, with negligible contribution from Sb states. 
Consequently, the CP density provides a direct real-space picture of the V $3d$ local bonding structure. Compared with the symmetric high-density clouds in KS-DFT, Coulomb repulsion in CP-DFT strongly reshapes the local environment, flattening the conditional density near the reference electron and disrupting the $3d$ orbital symmetry (Fig.~\ref{fig:Kagome}b). The corresponding $a$-axis linecut in Fig.~\ref{fig:Kagome}c further shows the pronounced depletion around the reference electron.

\begin{figure}[htbp]
\centering
\includegraphics[width=0.47\textwidth]{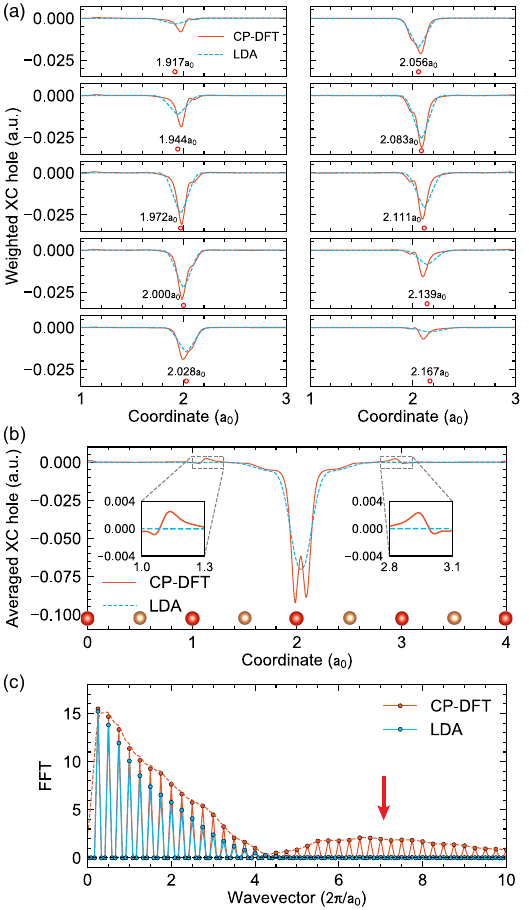}
\caption{(a) XC holes ($\lambda=1.0$) at 10 reference points (red circles, ranging from 1.917 $a_0$ to 2.167 $a_0$), weighted and normalized by the KS density, with \textit{y}=2.083 $a_0$. (b) XC holes along the $a$ axis in the supercell, averaged over 32 reference points in 1.61 $a_0$ $\sim$ 2.47 $a_0$, with linecut shown in Fig.~\ref{fig:Kagome}(b). (c) Fourier transform spectrum for the averaged XC hole in (b), with approximate envelope shown by dashed line. The wavevector unit is $\rm Q_{ \text{Bragg}}=2\pi/a_0$.}
\label{fig:xchole}
\end{figure}

The fine structure of the CP density offers deep insights into the intrinsic properties of these materials. Along the density linecut (Fig.~\ref{fig:Kagome}b), we analyze the fine-structure differences manifested in the XC hole (Fig.~\ref{fig:xchole}). For comparison, we calculate the LDA XC hole~\cite{Perdew1992a} (Supplementary Note~3 and Fig. S15~\cite{SuppMat}), which inherently exhibits a monotonic and spherical profile. Both are weighted and normalized by their respective KS densities. As the reference point is scanned left to right across the $d$-orbital density peaks (1.917 $a_0$ to 2.167 $a_0$), the XC holes initially intensify and then diminish (Fig.~\ref{fig:xchole}a). For each reference point, the CP-DFT XC hole exhibits an adjacent main peak and a secondary shoulder or peak near a neighboring $d$-orbital KS density peak, while the LDA hole displays only a single featureless peak. These distinctions are further emphasized in the averaged XC hole, calculated over 32 reference points (1.61 $a_0$ to 2.47 $a_0$). The CP-DFT XC hole is heavily concentrated in the on-top region (near 2.1 $a_0$), whereas the LDA result remains broad and diffuse. This double-peak structure between 1.9 $a_0$ and 2.1 $a_0$ reveals local deformation of the V $3d$ electrons, and is further verified by linecuts along another $a$-axis direction and the $b$-axis direction (Supplementary Fig. S9~\cite{SuppMat}).

A possible concern is dependence on the choice of approximation for the KS calculations (here PBE) via the self-consistent densities. Such density differences are typically very small and only significant when inserted inside different XC functionals~\cite{KSB13,KSSBS26}. Because CP-DFT uses the KS scheme to directly calculate CP densities, such differences should be negligible on our scales. To check, we performed an LDA-based CP-DFT control calculation using the same reference points and density line (Supplementary Fig. S10~\cite{SuppMat}), yielding no visible effect on the CP-DFT XC-hole.

 The Coulomb interaction induces small positive density waves near the V atom (at both 1 $a_0$ and 3 $a_0$), yielding an enhanced probability of finding two electrons at a spatial separation of $\sim1\,a_0$, as highlighted in the inset figures and in the scaled XC holes shown in Supplementary Fig. S11~\cite{SuppMat}. These alternating positive and negative peaks are responsible for the local CDW-related modulation, contributing to the amplification and reduction of the CDW-related amplitude. This counterintuitive feature—where strong repulsion coincides with an enhanced probability of finding two electrons at a finite separation—suggests the emergence of non-local correlation patterns within the pair density, and is starkly visible in the Fourier transform of the density linecut (Fig.~\ref{fig:xchole}c). While the LDA result decays rapidly in momentum space, the CP-DFT results exhibit prominent peaks between 5 $\rm Q_{\text{Bragg}}$ and 8 $\rm Q_{\text{Bragg}}$ arising from the double-peak structure, along with a small enhancement at Q=1/2 $\rm Q_{\text{Bragg}}$. The high-frequency peaks imply short-range order that is missed in standard DFT, underscoring the complex electron-electron interactions in the Kagome lattice.

\begin{figure}[htbp]
\centering
\includegraphics[width=0.47\textwidth]{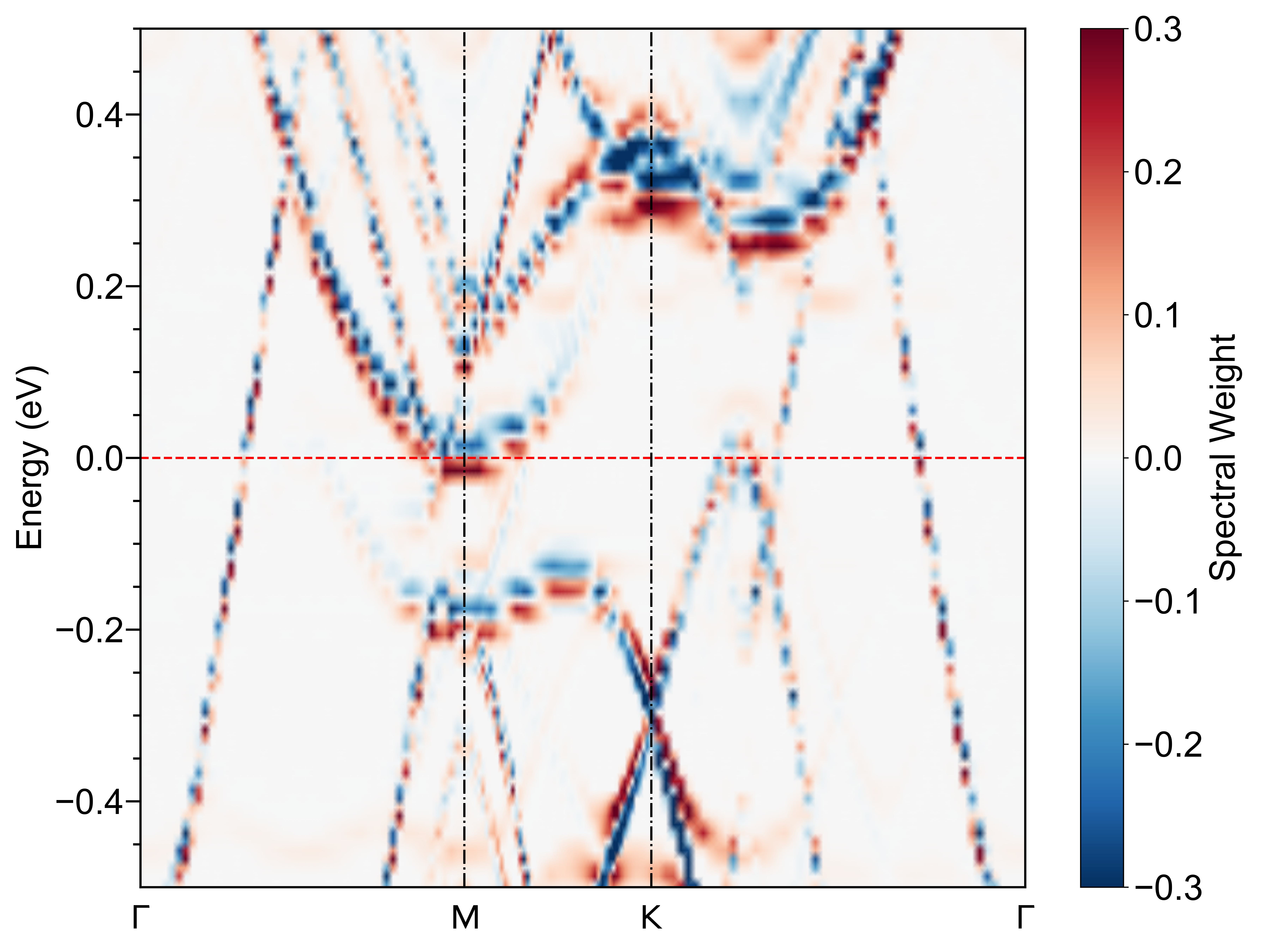}
\caption{Difference in spectral weight between $\lambda=1.0$ and $\lambda=0.0$ along $\Gamma-M-K-\Gamma$ high-symmetry path. The Fermi level is indicated by the red dashed line. The reference point is located at $x=2.083~a_0, y=2.083~a_0$ shown in Fig. \ref{fig:Kagome}.}
\label{fig:dos}
\end{figure}

Finally, we examine how these electron-electron interactions affect the momentum-space electronic structure. Fixing a reference electron breaks translational symmetry, so we diagonalize the CP Hamiltonian in the supercell and unfold the resulting eigenstates onto the primitive Brillouin zone to obtain the spectral weight (Supplementary Fig.~S12~\cite{SuppMat}). The difference between $\lambda=1$ and $\lambda=0$ isolates the correlation-induced changes (Fig.~\ref{fig:dos}). 
At the M point, the upper mixture-type van Hove singularity (vHS) at 0.1~eV~\cite{Kang2022} remains unperturbed. This preserves the band dispersion's charge order gap, implying that electron-electron correlation does not destroy the experimentally observed charge order. More importantly, CP-DFT predicts a 20--40~meV downward shift in both the energy and spectral weight of the vHS band near the Fermi level, suggesting correlation‑driven band shifts — similar in spirit to those observed in CsCr$_3$Sb$_5$, where standard DFT fails without phenomenological correlation~\cite{Wang2025Spin,Xie2025}. Such a prediction emerges self-consistently from $d$-orbital correlations and awaits further experimental verification. This downward shift is associated with a negative component of the CP effective potential. Far from the reference point (Supplementary Fig. S13~\cite{SuppMat}), the XC-hole-induced negative potential screens the bare Coulomb repulsion and locally yields a net negative potential. These spectral changes suggest that electron-electron interactions, properly captured by the correlated pair density, influence the low-energy electronic landscape.

We implemented CP-DFT for solids and computed the pointwise pair density and XC hole. Benchmark calculations for He, Na and Si yield XC energies close to standard KS results. In hydrogen chains, CP-DFT captures a long-range, approximately power-law decay of the XC hole. In the Kagome metal CsV$_3$Sb$_5$, CP-DFT reveals correlated structures with local bonding information of $3d$ orbitals. These correlation-induced modulations show a slightly enhanced CDW-related feature, providing a real-space view of how electron-electron interactions reshape density correlations in a topological metal. Beyond conventional energy-based density functional approaches, periodic CP-DFT opens a route to analyzing the real-space correlation patterns that underlie electronic instabilities and ordering phenomena, and may link such correlations to collective electronic phases, such as superconductivity or charge order.

\begin{acknowledgments}
E.K.U. Gross acknowledges financial support from the Fondation de l'Ecole Polytechnique within the Gaspard Monge programme. K. Burke is supported by NSF grant DMR-2427903. P. You thanks Sheng Meng for discussions.
\end{acknowledgments}
\nocite{Kato1957,Sun2015,Kittel1986,Schouten2025PRB,
Umrigar1994,PribramJones2015}
\bibliography{references}

@article{McCarty2020,
  title = {Bypassing the Energy Functional in Density Functional Theory: Direct Calculation of Electronic Energies from Conditional Probability Densities},
  author = {McCarty, R. J. and Perchak, D. and Pederson, R. and Evans, R. and Qiu, Y. and White, S. R. and Burke, K.},
  journal = {Phys. Rev. Lett.},
  volume = {125},
  pages = {266401},
  year = {2020}
}

@article{Pederson2022,
  title = {Conditional probability density functional theory},
  author = {Pederson, R. and Chen, J. and White, S. R. and Burke, K.},
  journal = {Phys. Rev. B},
  volume = {105},
  pages = {245138},
  year = {2022}
}

@article{Agapito2015,
  title = {Reformulation of {DFT+U} as a Pseudohybrid Hubbard Density Functional for Accelerated Materials Discovery},
  author = {Agapito, L. A. and Curtarolo, S. and Buongiorno Nardelli, M.},
  journal = {Phys. Rev. X},
  volume = {5},
  pages = {011006},
  year = {2015}
}

@article{Sharma2013,
  title = {Spectral density and metal-insulator phase transition in Mott insulators within reduced density matrix functional theory},
  author = {Sharma, S. and Dewhurst, J. K. and Shallcross, S. and Gross, E. K. U.},
  journal = {Phys. Rev. Lett.},
  volume = {110},
  pages = {116403},
  year = {2013}
}

@article{Zhu2015,
  title = {Classification of charge density waves based on their nature},
  author = {Zhu, X. and Cao, Y. and Zhang, J. and Plummer, E. W. and Guo, J.},
  journal = {Proc. Natl. Acad. Sci. U.S.A.},
  volume = {112},
  pages = {2367-2371},
  year = {2015}
}

@article{Chen2021,
  title = {Roton pair density wave in a strong-coupling kagome superconductor},
  author = {Chen, H. and Yang, H. and Hu, B. and Zhao, Z. and Yuan, J. and Xing, Y. and Qian, G. and Huang, Z. and Li, G. and Ye, Y. and Ma, S. and Ni, S. and Zhang, H. and Yin, Q. and Gong, C. and Tu, Z. and Lei, H. and Tan, H. and Zhou, S. and Shen, C. and Dong, X. and Yan, B. and Wang, Z. and Gao, H. J.},
  journal = {Nature},
  volume = {599},
  pages = {222-228},
  year = {2021}
}

@article{Tan2021,
  title = {Charge density waves and electronic properties of superconducting kagome metals},
  author = {Tan, H. and Liu, Y. and Wang, Z. and Yan, B.},
  journal = {Phys. Rev. Lett.},
  volume = {127},
  pages = {046401},
  year = {2021}
}

@article{Song2023,
  title = {Anomalous enhancement of charge density wave in kagome superconductor {CsV$_3$Sb$_5$} approaching the 2D limit},
  author = {Song, B. and Ying, T. and Wu, X. and Xia, W. and Yin, Q. and Zhang, Q. and Song, Y. and Yang, X. and Guo, J. and Gu, L. and Chen, X. and Hu, J. and Schnyder, A. P. and Lei, H. and Guo, Y. and Li, S.},
  journal = {Nat. Commun.},
  volume = {14},
  pages = {2492},
  year = {2023}
}

@article{Perdew1992a,
  title = {Pair-distribution function and its coupling-constant average for the spin-polarized electron gas},
  author = {Perdew, J. P. and Wang, Y.},
  journal = {Phys. Rev. B},
  volume = {46},
  pages = {12947-12954},
  year = {1992}
}

@article{GoriGiorgi2005,
  title = {Simple model for the spherically and system-averaged pair density: Results for two-electron atoms},
  author = {Gori-Giorgi, P. and Savin, A.},
  journal = {Phys. Rev. A},
  volume = {71},
  pages = {032513},
  year = {2005}
}

@article{GoriGiorgi2007,
  title = {Kohn-Sham calculations combined with an average pair-density functional theory},
  author = {Gori-Giorgi, P. and Savin, A.},
  journal = {Int. J. Mod. Phys. B},
  volume = {21},
  pages = {2449-2459},
  year = {2007}
}

@article{LiManni2014,
  title = {Multiconfiguration Pair-Density Functional Theory},
  author = {Li Manni, G. and Carlson, R. K. and Luo, S. and Ma, D. and Olsen, J. and Truhlar, D. G. and Gagliardi, L.},
  journal = {J. Chem. Theory Comput.},
  volume = {10},
  pages = {3669-3680},
  year = {2014}
}

@article{Kato1957,
  title = {On the eigenfunctions of many-particle systems in quantum mechanics},
  author = {Kato, T.},
  journal = {Commun. Pure Appl. Math.},
  volume = {10},
  pages = {151-177},
  year = {1957}
}

@book{Fiolhais2003,
  title = {A primer in density functional theory},
  author = {Fiolhais, C. and Nogueira, F. and Marques, M. A. L.},
  publisher = {Springer Science \& Business Media},
  year = {2003}
}

@article{Perdew1996,
  title = {Generalized gradient approximation made simple},
  author = {Perdew, J. P. and Burke, K. and Ernzerhof, M.},
  journal = {Phys. Rev. Lett.},
  volume = {77},
  pages = {3865-3868},
  year = {1996}
}

@article{Sun2015,
  title = {Strongly Constrained and Appropriately Normed Semilocal Density Functional},
  author = {Sun, J. and Ruzsinszky, A. and Perdew, J. P.},
  journal = {Phys. Rev. Lett.},
  volume = {115},
  pages = {036402},
  year = {2015}
}

@article{Soler2002,
  title = {The SIESTA method for ab initio order-N materials simulation},
  author = {Soler, J. M. and Artacho, E. and Gale, J. D. and Garc\'{i}a, A. and Junquera, J. and Ordej\'{o}n, P. and S\'{a}nchez-Portal, D.},
  journal = {J. Phys.: Condens. Matter},
  volume = {14},
  pages = {2745-2779},
  year = {2002}
}

@article{Garcia2020,
  title = {Siesta: Recent developments and applications},
  author = {Garcia, A. and Papior, N. and Akhtar, A. and Artacho, E. and Blum, V. and Bosoni, E. and Brandimarte, P. and Brandbyge, M. and Cerda, J. I. and Corsetti, F. and Cuadrado, R. and Dikan, V. and Ferrer, J. and Gale, J. and Garcia-Fernandez, P. and Garcia-Suarez, V. M. and Garcia, S. and Huhs, G. and Illera, S. and Korytar, R. and Koval, P. and Lebedeva, I. and Lin, L. and Lopez-Tarifa, P. and Mayo, S. G. and Mohr, P. and Ordejon, P. and Postnikov, A. and Pouillon, Y. and Pruneda, M. and Robles, R. and Sanchez-Portal, D. and Soler, J. M. and Ullah, R. and Yu, V. W. and Junquera, J.},
  journal = {J. Chem. Phys.},
  volume = {152},
  pages = {204108},
  year = {2020}
}

@article{Ortiz2020,
  title = {{CsV$_3$Sb$_5$}: A {Z}$_2$ topological kagome metal with a superconducting ground state},
  author = {Ortiz, B. R. and Teicher, S. M. L. and Hu, Y. and Zuo, J. L. and Sarte, P. M. and Schueller, E. C. and Abeykoon, A. M. M. and Krogstad, M. J. and Rosenkranz, S. and Osborn, R. and Seshadri, R. and Balents, L. and He, J. and Wilson, S. D.},
  journal = {Phys. Rev. Lett.},
  volume = {125},
  pages = {247002},
  year = {2020}
}

@article{Liang2021,
  title = {Three-dimensional charge density wave and surface-dependent vortex-core states in a kagome superconductor {CsV$_3$Sb$_5$}},
  author = {Liang, Z. and Hou, X. and Zhang, F. and Ma, W. and Wu, P. and Zhang, Z. and Yu, F. and Ying, J. J. and Jiang, K. and Shan, L. and Wang, Z. and Chen, X. H.},
  journal = {Phys. Rev. X},
  volume = {11},
  pages = {031026},
  year = {2021}
}

@article{Zhao2021,
  title = {Cascade of correlated electron states in the kagome superconductor {CsV$_3$Sb$_5$}},
  author = {Zhao, H. and Li, H. and Ortiz, B. R. and Teicher, S. M. L. and Park, T. and Ye, M. and Wang, Z. and Balents, L. and Wilson, S. D. and Zeljkovic, I.},
  journal = {Nature},
  volume = {599},
  pages = {216-221},
  year = {2021}
}

@article{Kang2022,
  title = {Twofold van Hove singularity and origin of charge order in topological kagome superconductor {CsV$_3$Sb$_5$}},
  author = {Kang, M. and Fang, S. and Kim, J.-K. and Ortiz, B. R. and Ryu, S. H. and Kim, J. and Yoo, J. and Sangiovanni, G. and Di Sante, D. and Park, B.-G. and Jozwiak, C. and Bostwick, A. and Rotenberg, E. and Kaxiras, E. and Wilson, S. D. and Park, J.-H. and Comin, R.},
  journal = {Nat. Phys.},
  volume = {18},
  pages = {301-308},
  year = {2022}
}

@misc{SuppMat,
title = {See {Supplemental Material} at [URL] for implementation details, calculation methods, pair-distribution functions, and additional benchmark and materials results, which includes {Refs.} [40--45].}
}

@article{Burke1998,
    author = {Burke, Kieron and Perdew, John P. and Ernzerhof, Matthias},
    title = {Why semilocal functionals work: Accuracy of the on-top pair density and importance of system averaging},
    journal = {The Journal of Chemical Physics},
    volume = {109},
    number = {10},
    pages = {3760-3771},
    year = {1998},
    month = {09},
    issn = {0021-9606},
}

@article{Jones2015,
  title = {Density functional theory: Its origins, rise to prominence, and future},
  author = {Jones, R. O.},
  journal = {Rev. Mod. Phys.},
  volume = {87},
  issue = {3},
  pages = {897--923},
  numpages = {27},
  year = {2015},
  month = {Aug},
  publisher = {American Physical Society},
}

@article{Lejaeghere01012014,
author = {K. Lejaeghere and V. Van Speybroeck and G. Van Oost and S. Cottenier},
title = {Error Estimates for Solid-State Density-Functional Theory Predictions: An Overview by Means of the Ground-State Elemental Crystals},
journal = {Critical Reviews in Solid State and Materials Sciences},
volume = {39},
number = {1},
pages = {1--24},
year = {2014},
publisher = {Taylor \& Francis},
}

@article{Kaplan2023,
   author = "Kaplan, Aaron D. and Levy, Mel and Perdew, John P.",
   title = "The Predictive Power of Exact Constraints and Appropriate Norms in Density Functional Theory", 
   journal= "Annual Review of Physical Chemistry",
   year = "2023",
   volume = "74",
   number = "Volume 74, 2023",
   pages = "193-218",
   publisher = "Annual Reviews",
   issn = "1545-1593",
   type = "Journal Article",

  }

@article{BurkePerdew1995,
author = {Burke, Kieron and Perdew, John P.},
title = {Real-space analysis of the exchange-correlation energy},
journal = {International Journal of Quantum Chemistry},
volume = {56},
number = {4},
pages = {199-210},
year = {1995},
}

@article{PerdewSavin1995,
  title = {Escaping the symmetry dilemma through a pair-density interpretation of spin-density functional theory},
  author = {Perdew, John P. and Savin, Andreas and Burke, Kieron},
  journal = {Phys. Rev. A},
  volume = {51},
  issue = {6},
  pages = {4531--4541},
  numpages = {0},
  year = {1995},
  month = {Jun},
  publisher = {American Physical Society},
}

@article{JonesGunnarsson1989,
  title = {The density functional formalism, its applications and prospects},
  author = {Jones, R. O. and Gunnarsson, O.},
  journal = {Rev. Mod. Phys.},
  volume = {61},
  issue = {3},
  pages = {689--746},
  numpages = {0},
  year = {1989},
  month = {Jul},
  publisher = {American Physical Society},
}

@article{Perchak2022,
  title = {Correlation energy of the uniform electron gas determined by ground-state conditional probability density functional theory},
  author = {Perchak, Dennis and McCarty, Ryan J. and Burke, Kieron},
  journal = {Phys. Rev. B},
  volume = {105},
  issue = {16},
  pages = {165143},
  numpages = {13},
  year = {2022},
  month = {Apr},
  publisher = {American Physical Society},
}

@article{LangrethPerdew1975,
title = {The exchange-correlation energy of a metallic surface},
journal = {Solid State Communications},
volume = {17},
number = {11},
pages = {1425-1429},
year = {1975},
issn = {0038-1098},
author = {D.C. Langreth and J.P. Perdew},
}

@article{Gorling1993,
  title = {Correlation-energy functional and its high-density limit obtained from a coupling-constant perturbation expansion},
  author = {G\"orling, Andreas and Levy, Mel},
  journal = {Phys. Rev. B},
  volume = {47},
  issue = {20},
  pages = {13105--13113},
  numpages = {0},
  year = {1993},
  month = {May},
  publisher = {American Physical Society},
}

@article{Wang2025Spin,
  title={Spin excitations and flat electronic bands in a Cr-based kagome superconductor},
  author={Wang, Zehao and Guo, Yucheng and Huang, Hsiao-Yu and Xie, Fang and others},
  journal={Nat. Commun.},
  volume={16},
  pages={7573}, 
  year={2025},
  publisher={Nature Publishing Group}
}

@article{Xie2025,
  title = {Electron correlations in the kagome flat band metal ${\mathrm{CsCr}}_{3}{\mathrm{Sb}}_{5}$},
  author = {Xie, Fang and Fang, Yuan and Li, Ying and Huang, Yuefei and Chen, Lei and Setty, Chandan and Sur, Shouvik and Yakobson, Boris and Valent\'{\i}, Roser and Si, Qimiao},
  journal = {Phys. Rev. Res.},
  volume = {7},
  issue = {2},
  pages = {L022061},
  numpages = {7},
  year = {2025},
  month = {Jun},
  publisher = {American Physical Society},
}

@article{Motta2020Hydrogen,
  title={Ground-State Properties of the Hydrogen Chain: Dimerization, Insulator-to-Metal Transition, and Magnetic Phases},
  author={Motta, Mario and Genovese, Claudio and Ma, Fengjie and Cui, Zhi-Hao and Sawaya, Randy and Chan, Garnet Kin-Lic and Chepiga, Natalia and Helms, Phillip and Jimenez-Hoyos, Carlos A. and Millis, Andrew J. and Ray, Ushnish and Ronca, Enrico and Shi, Hao and Sorella, Sandro and Stoudenmire, Edwin M. and White, Steven R. and Zhang, Shiwei},
  journal={Phys. Rev. X},
  volume={10},
  pages={031058},
  year={2020},
  publisher={American Physical Society},
}

@article{Motta2017,
  title = {Towards the Solution of the Many-Electron Problem in Real Materials: Equation of State of the Hydrogen Chain with State-of-the-Art Many-Body Methods},
  author = {Motta, Mario and Ceperley, David M. and Chan, Garnet Kin-Lic and Gomez, John A. and Gull, Emanuel and Guo, Sheng and Jim\'enez-Hoyos, Carlos A. and Lan, Tran Nguyen and Li, Jia and Ma, Fengjie and Millis, Andrew J. and Prokof'ev, Nikolay V. and Ray, Ushnish and Scuseria, Gustavo E. and Sorella, Sandro and Stoudenmire, Edwin M. and Sun, Qiming and Tupitsyn, Igor S. and White, Steven R. and Zgid, Dominika and Zhang, Shiwei},
  collaboration = {Simons Collaboration on the Many-Electron Problem},
  journal = {Phys. Rev. X},
  volume = {7},
  issue = {3},
  pages = {031059},
  numpages = {28},
  year = {2017},
  month = {Sep},
  publisher = {American Physical Society},
}

@article{Stoudenmire2012,
  title={One-Dimensional Continuum Electronic Structure with the Density-Matrix Renormalization Group and Its Implications for Density-Functional Theory},
  author={Stoudenmire, E. M. and Wagner, Lucas O. and White, Steven R. and Burke, Kieron},
  journal={Phys. Rev. Lett.},
  volume={109},
  pages={056402},
  year={2012},
}

@article{KSSBS26,
Pub-num = {242},
Author = {Youngsam Kim and Suhwan Song and Mihira Sogal and Kieron Burke and Eunji Sim},
Title = {Density-Corrected Density Functional Theory for Solids},
Year = {2026},
Journal = {Phys. Rev. Lett., To Appear},
Doi = {https://doi.org/10.26434/chemrxiv.15002348/v1},
Supp-info = {KSSBS26s}
}

@article{KSB13,
	Pub-num = {141},
	Author = {Min-Cheol Kim and Eunji Sim and Kieron Burke},
	journal = {Phys. Rev. Lett.},
	Title = {Understanding and reducing errors in density functional calculations},
	volume = {111},
	issue = {7},
	pages = {073003},
	year = {2013},
	month = {Aug},
	supp-info = {KSB13_supp}
}

@article{Liu2020Hydrogen,
  title={Unveiling the Finite Temperature Physics of Hydrogen Chains via Auxiliary Field Quantum Monte Carlo},
  author={Liu, Yuan and Shen, Tong and Zhang, Hang and Rubenstein, Brenda},
  journal={J. Chem. Theory Comput.},
  volume={16},
  number={7},
  pages={4298--4314},
  year={2020},
}

@book{Kittel1986,
  title = {Introduction to Solid State Physics},
  author = {Kittel, C.},
  publisher = {John Wiley \& Sons},
  year = {1986}
}

@article{Schouten2025PRB,
  title = {Two-electron reduced density matrix method for the electronic structure of correlated materials},
  author = {Schouten, A. O. and Ewing, S. and Mazziotti, D. A.},
  journal = {Phys. Rev. B},
  volume = {112},
  pages = {235130},
  year = {2025}
}

@article{Umrigar1994,
  title = {Accurate exchange-correlation potentials and total-energy components for the helium isoelectronic series},
  author = {Umrigar, C. J. and Gonze, X.},
  journal = {Phys. Rev. A},
  volume = {50},
  pages = {3827},
  year = {1994}
}

@article{PribramJones2015,
  title = {DFT: A Theory Full of Holes?},
  author = {Pribram-Jones, A. and Gross, D. A. and Burke, K.},
  journal = {Annu. Rev. Phys. Chem.},
  volume = {66},
  pages = {283},
  year = {2015}
}

\end{document}